\documentclass[aps,prl,twocolumn,superscriptaddress]{revtex4-1} 
\pdfoutput=1
\usepackage{float}
\usepackage{graphicx}
\usepackage{xcolor}
\usepackage{soul}
\usepackage{amsmath}
\usepackage{amsfonts}
\usepackage{bbm}
\usepackage{xcolor}
\usepackage{enumitem}
\usepackage{svg}
\usepackage{url}

\begin{document}


\title{First principles study of ferroelastic twins in halide perovskites}


\author{Andrew R. Warwick}
\email[]{andrew.warwick11@imperial.ac.uk}

\affiliation{Department of Materials, Imperial College London, London SW7 2AZ, United Kingdom }

\author{Jorge \'{I}\~{n}iguez}
\affiliation{Materials Research and Technology Department, Luxembourg Institute of Science and Technology, 5 avenue des Hauts-Fourneaux, 4362 Esch/Alzette, Luxembourg}
\affiliation{Physics and Materials Science Research Unit, University of Luxembourg, 41 Rue du Brill, L-4422 Belvaux, Luxembourg}

\author{Peter D. Haynes}
\affiliation{Department of Materials, Imperial College London, London SW7 2AZ, United Kingdom }

\author{Nicholas C. Bristowe}
\affiliation{School of Physical Sciences, University of Kent, Canterbury,CT2 7NH, United Kingdom}
\affiliation{Department of Materials, Imperial College London, London SW7 2AZ, United Kingdom }


\date{\today}

\begin{abstract}
	
	We present an \textit{ab initio} simulation of $90^{\circ}$ ferroelastic twins that were recently observed in methyl ammonium lead iodide \cite{RothmannLiZhuEtAl2017}. %
	There are two inequivalent types of $90^{\circ}$ walls that we calculate to act as either electron or hole sinks which suggests a possible route to enhancing charge carrier separation in photovoltaic devices. %
	Despite separating non-polar domains, we show these walls to have a substantial in-plane polarisation of $\sim 6 \phantom{|} \mu \text{C}\phantom{|}\text{cm}^{-2}$, due in part to flexoelectricity. %
	We suggest this in turn could allow for the photoferroic effect and create efficient pathways for photocurrents within the wall. %
	
\end{abstract}


\maketitle


\section{Introduction \label{sec:intro}}
	
	The synthesis of a hybrid organic-inorganic halide pervoskite (`HOIP') solar cell was first reported in 2009 \cite{KojimaTeshimaShiraiEtAl2009}. %
	Since then, the power conversion efficiencies (PCE) of these devices have climbed rapidly from 14.3\% to 23.7\% \cite{NRELchart}. %
	A wide range of materials, in particular methylammonium lead iodide ($\mathrm{CH_3 NH_3 PbI_3}$, `MAPI'), have been considered to investigate this class of photovoltaic devices. %
	Key issues regarding the stability and efficiency of these materials are an ongoing area of research (\textit{cf.}\ reviews \cite{Bhatt2017review, Huang2017NatReview} and references therein), not least of which is the origin of their high PCE. %
	In this context, one important issue still undergoing investigation is whether MAPI is ferroelectric and, by extension, how the photoferroic effect \cite{fridkin1979photoferroelectrics, sturman1992photovoltaic, ButlerEtAl2015} might play a role in the reported PCEs. %
	As of yet, there is no clear consensus on the former issue \cite{FrostEtAl2014, KutesEtAl2014, FanEtAl2015, SharadaEtAl2016, KimEtAl2016, RakitaEtAl2017, HoqueEtAl2016}. %
	In addition to multiple related studies on the bulk material, there have been a number of publications on planar defects such as grain boundaries and domain walls which we briefly summarise here (more detailed reviews can be found in \cite{WhalleyEtAl2017, LiuEtAlTwinRev2018}). %
	
	At a defect boundary, properties not present in the bulk may emerge at the interface itself, chiefly due to the local structural distortion. %
	For instance, a bulk property may change orientation across the boundary and thus some components will locally vanish or be enhanced. %
	Hence, physics that is ordinarily suppressed in the bulk may manifest at the wall. %
	Theoretical studies on grain boundaries \cite{ShanSaidi2017GB, WanJianEtAl2014GB, McKenna2018faGB, GuoEtAl2017GB, WanJianEtAl2015GB} and ferroelectric domain walls \cite{LiuZhengKoocherEtAl2015, ChenEtAl2018FEDW,FrostEtAl2014} have proposed a variety of ways in which these types of planar defects may influence the electronic properties of HOIPs. %
	Some first--principles simulations suggest that the presence of grain boundaries is not detrimental to photovoltaic performance \cite{GuoEtAl2017GB, WanJianEtAl2015GB, WanJianEtAl2014GB}. %
	Conversely, experimental evidence indicates charge carrier recombination is enhanced at these defects \cite{BischakEtAl2015GB}. %
	It has been proposed that ferroelectric domain walls (`twins') can be formed via interfacing polar domains in which the methyl ammonium (MA) molecular dipoles are aligned. %
	Calculations on these twins indicate beneficial effects such as locally diminished band gaps \cite{LiuZhengKoocherEtAl2015, ChenEtAl2018FEDW} or enhanced charge carrier diffusion lengths \cite{FrostEtAl2014}. %
	
	Recently, ferro\textit{elastic} twins in MAPI were observed at room temperature and their structure has been determined \cite{RothmannLiZhuEtAl2017, LiuEtAl2018feTwinAPL, LiuEtAl2018feTwinNM}. %
	We emphasise that these types of defect are distinct from the aforementioned grain boundaries and ferroelectric twins. %
	Ferroelastic domain walls separate domains with different strain states. %
	Unlike a general grain boundary, the domains are `crystallographically' related \cite{ITD_C3} and, in contrast to ferroelectric twins, these domains can be non-polar (which may be the case in MAPI). %
	Nonetheless, to the best of our knowledge, an \textit{ab initio} simulation of the ferroelastic twins observed and described here has not been carried out. %
	
	Here, we report a density-functional theory simulation of the twins characterised in \cite{RothmannLiZhuEtAl2017, LiuEtAl2018feTwinAPL, LiuEtAl2018feTwinNM}. 
	We find the walls to be thin, polar and have a low formation energy. %
	The effect of these properties on the photovoltaic performance of halide perovkistes is discussed in addition to possible avenues for domain-wall engineering such devices. %

\section{Methods \& computational details}
\label{sec:met}

	\begin{figure}[h]
		\centering
		\includegraphics[trim={0 0 0 0},clip,width=0.5\textwidth,angle=0]{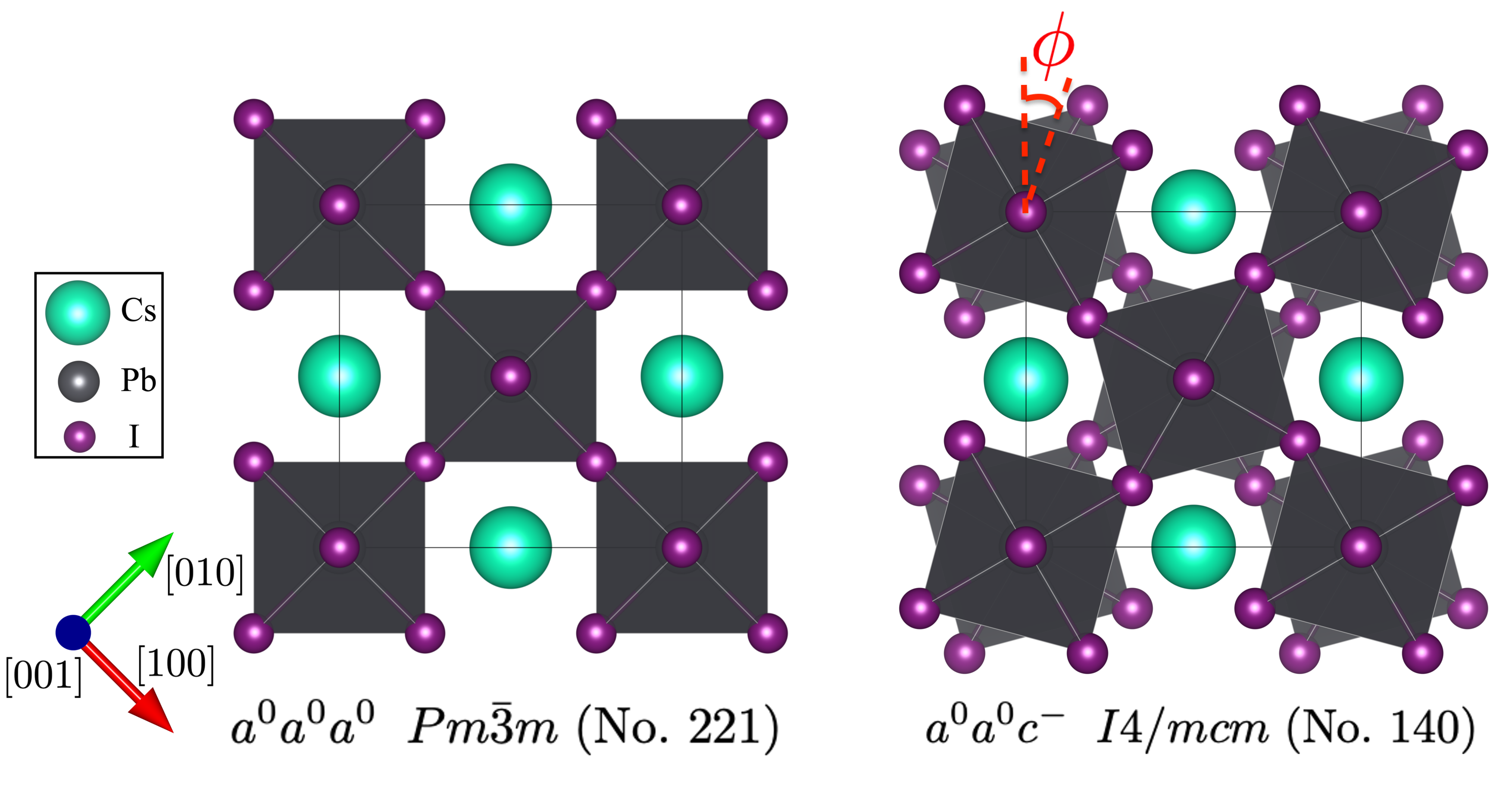}
		\caption{Illustration of the cubic $Pm\bar{3}m$ and tetragonal $I4/mcm$ phases of CsPbI\textsubscript{3}. The $I4/mcm$ phase is characterised by octahedral tilting about one axis. The octahedra maintain corner connectivity and tilt about the [001] direction (pointing out of the page) through an angle $\phi$ in opposite senses along (001) planes as denoted by Glazer's notation $ a^{0}a^{0}c^{-} $. }
		\label{fig:CsPbIi4mcm}
	\end{figure}

	There are obvious technical challenges to account for the dynamic ordering of MA molecules in a zero Kelvin calculation. %
	Hence we studied these walls in $\mathrm{CsPbI_3}$, the Cs atom having a similar effective size to the MA molecule \cite{LeeEtAl2016}. %
	Furthemore, the valence and conduction band edges are known to be mainly dominated by Pb and I states \cite{BrivioWalkerWalsh2013}. %
	Hence, we expect MAPI twins to display the properties presented here with additional physics arising from the organic molecule. %
	In addition, the use of $\mathrm{CsPbI_3}$ in synthesising photovoltaic devices \cite{EperonPaternoSuttonEtAl2015} and doping MAPI with Cs to improve PCE \cite{ChoiJeongKimEtAl2014} make this topical in its own right. 

	MAPI and $\mathrm{CsPbI_3}$ adopt the perovskite structure of corner-sharing $\mathrm{I_6}$ octahedra with MA\textsuperscript{+}/Cs\textsuperscript{+} and Pb\textsuperscript{2+} cubic sub--lattices. %
	We studied the room temperature tetragonal phase of MAPI with $\mathrm{CsPbI_3}$ where the $\mathrm{I_6}$ octahedra tilt in antiphase with angle $\phi$ along one tetrad axis (shown in Fig. \ref{fig:CsPbIi4mcm}). 
	The phase has space group $I4/mcm$ (No.\ 140) \cite{WellerWeberHenryEtAl2015} with Glazer tilt pattern $a^0 a^0 c^-$ (indicating no octahedral tilts about [100] and [010] directions and an antiphase tilt about the [001] axis \cite{Glazer1972}). %
	This single domain state may be characterised by a pseudovector $\boldsymbol{\phi}$ parallel to the tilt axis and whose magnitude is equal to the tilt angle $\phi$. %
	Note that the symmetry of this anti-phase tilt makes the $I4/mcm$ space group necessarily non-polar. %
	Due to the spontaneous strain induced by $\phi$, such a state is termed improper ferroelastic. %
	
\begin{figure*}
		\centering
		\includegraphics[trim={0 0 0 0},clip,width=1\textwidth,angle=0]{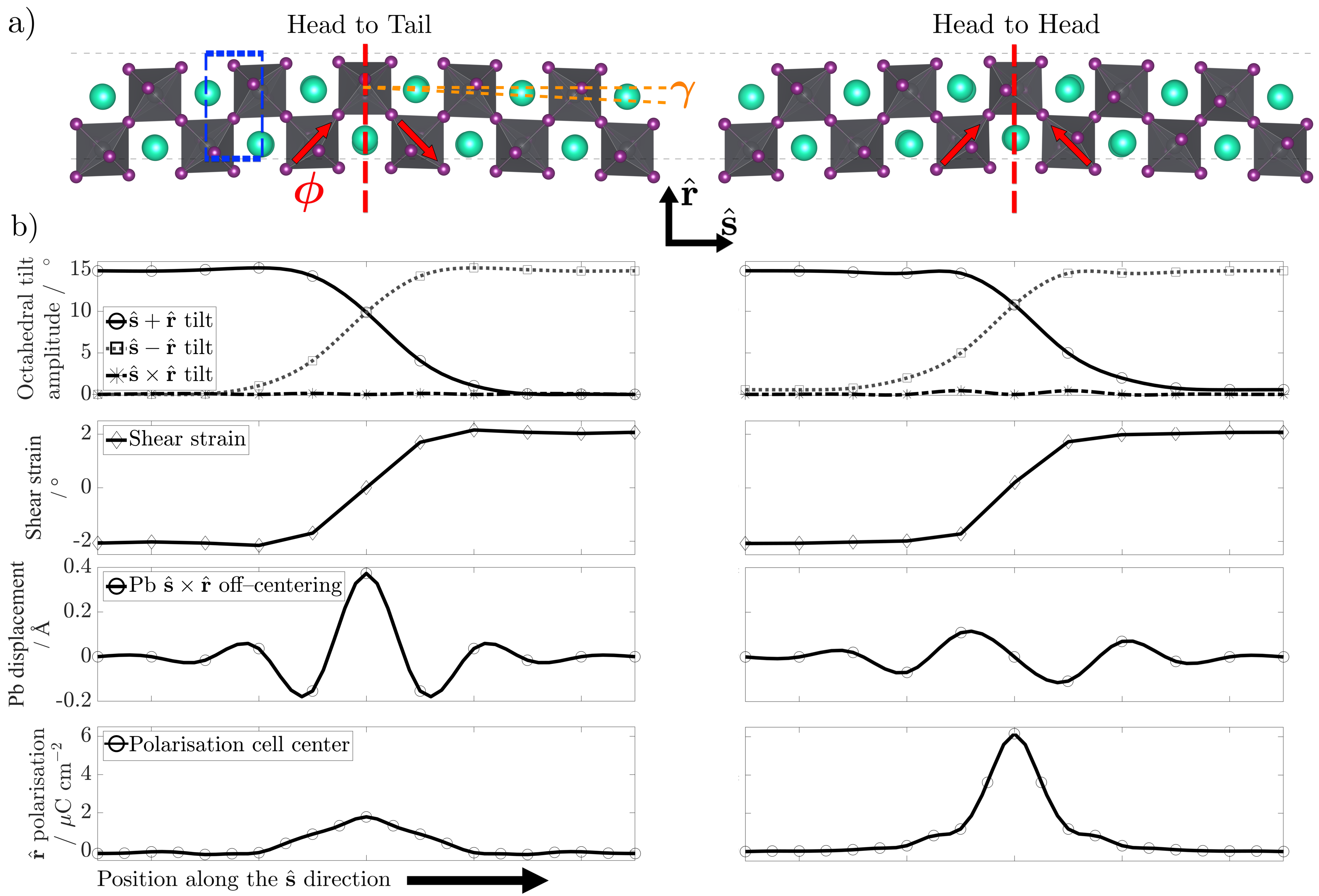}
		\caption{ a) Structure and b) properties of the HT (left) and HH (right) twins. %
			The plots are (from top to bottom) profiles of tilt amplitude, an estimate of shear strain, Pb\textsuperscript{2+} cation $\hat{\mathbf{s}}\times\hat{\mathbf{r}}$ off--centerings and the in--plane $\hat{\mathbf{r}}$ polarisation component (as obtained from the calculated displacements and Born effective charges (BEC)). %
			For both walls, the \textit{x}--axis marks units of distance along $\hat{\mathbf{s}}$ and is plotted to scale with the respective atomic configuration above. Tilt pseudovectors and the wall central plane are marked by red arrows (labelled $\boldsymbol{\phi}$) and a vertical dashed line in both structural schematics. Shear strain was quantified by the angle $\gamma$ coloured orange in a). Pb~off-centerings are plotted for octhedra visible in the plane of this page; for the adjacent plane of octhedra below/above along the $\hat{\mathbf{s}}\times\hat{\mathbf{r}}$ direction, the Pb off-centerings are in antiphase. The polarisations were calculated at cells indicated by the dashed blue box in the HT schematic.}
		\label{fig:structtiltpol}
	\end{figure*}

	The twins characterised by Rothmann \textit{et al}.\ \cite{RothmannLiZhuEtAl2017} are planar interfaces joining two domains whose tilt axes meet at $90^{\circ}$. %
	Figure \ref{fig:structtiltpol}a) shows two atomic structures consistent with these observed twins. 
	Directions perpendicular and parallel to the planar interface are labelled by unit vectors $\hat{\mathbf{s}}$ and $\hat{\mathbf{r}}$ respectively. %
	In fact, it can be shown that these are the only mechanically compatible $90^{\circ}$ twins \cite{ITD_C3} i.e.\ interfaces that do not introduce additional strains and stresses to this system. %
	Both twins in Fig.\ \ref{fig:structtiltpol}  may be characterised by the relative orientation of tilt pseudovectors across the wall's center. %
	Following standard terminology \cite{SchiaffinoStengel2017}, these are called `head to tail' (HT) or `head to head' (HH). %


	The twins were simulated with periodic boundary conditions in VASP 5.4.4 using the PBEsol exchange-correlation functional \cite{Perdew2008PBEsol}. %
	Valence electron configurations 6\textit{s}\textsuperscript{1}, 6\textit{s}\textsuperscript{2}6\textit{p}\textsuperscript{2} and 5\textit{s}\textsuperscript{2}5\textit{p}\textsuperscript{5} were employed for Cs, Pb and I respectively with the supplied Projector-Augmented-Wave pseudopotentials generated in 2002 \cite{Bloechl1994,KresseJoubert1999,KresseHafner1993,KresseFurth1996}. %
	We used supercells containing 200 atoms, the centers of which are shown in Fig.\ \ref{fig:structtiltpol}a), such that the cell contained two walls (one at the center and another shared at the edges) yet was sufficiently large to recover the bulk properties between both walls. %
	With an energy cutoff of 500 eV for the plane-wave basis and a $4\times5\times1$ Monkhorst-Pack grid \cite{MonkhorstPack1976}, both supercells were relaxed such that the maximum force on any atom is less than 10~meV~\AA\textsuperscript{-1}. %

\section{Results \& discussion}
\label{sec:result}

	\begin{figure}[h]
		\centering
		\begin{tabular}{@{}c@{}}
			\hspace*{-0.6cm}\includegraphics[trim={0 0 0 0},clip,width=.5\textwidth,angle=0]{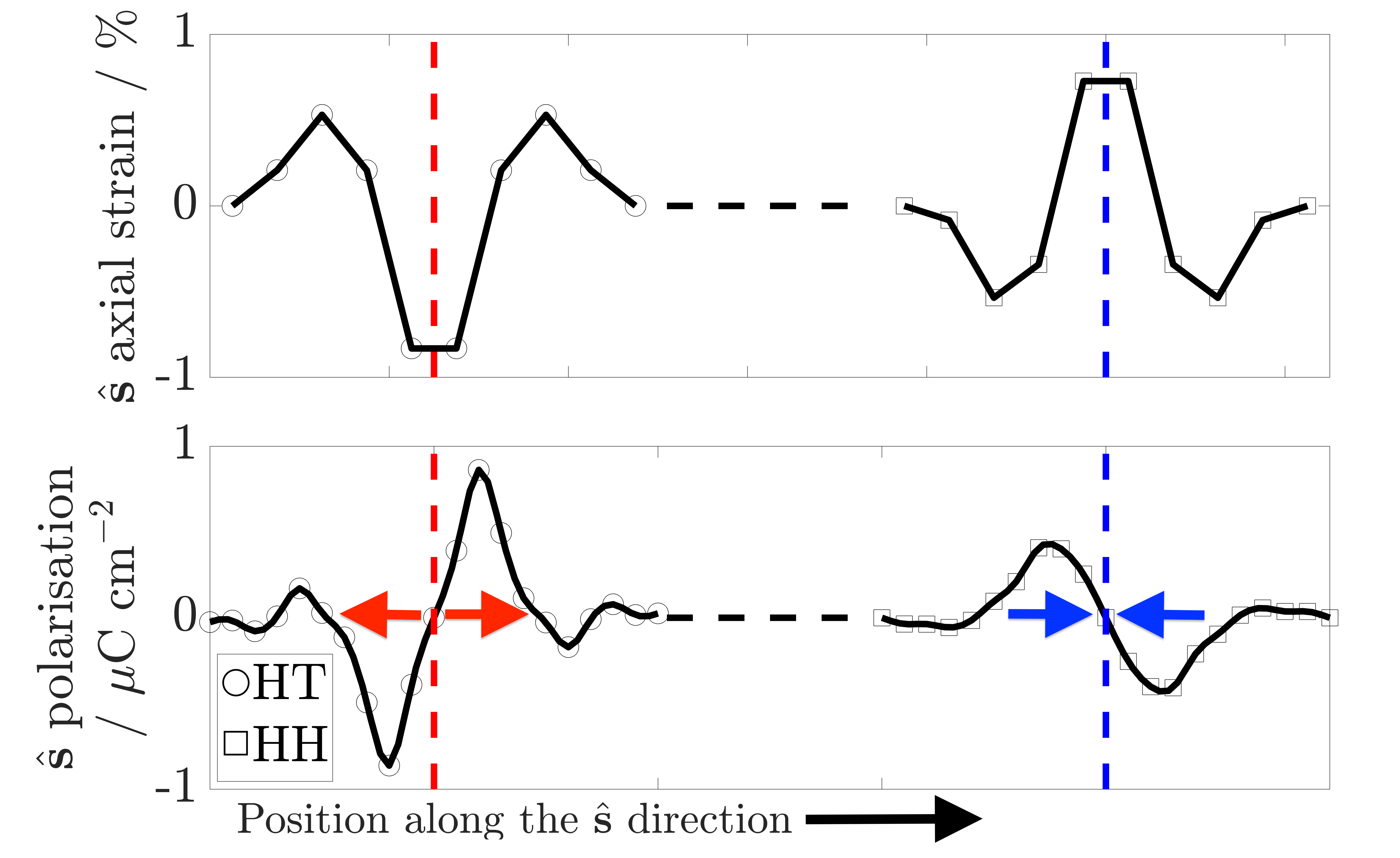} \\
			\vspace*{-0.6cm}\includegraphics[trim={0 0 0 0},clip,width=.5\textwidth,angle=0]{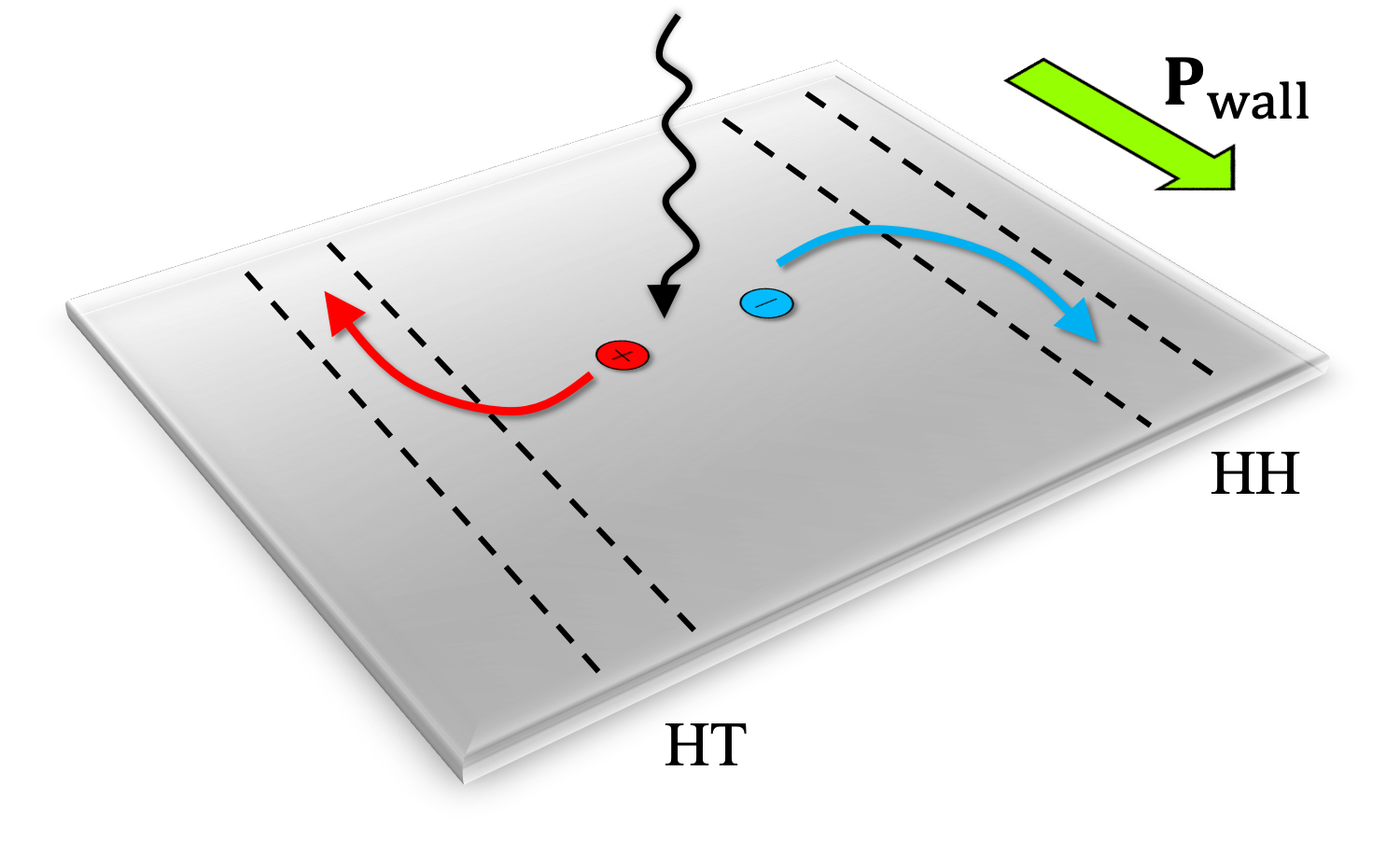}
		\end{tabular}
		\caption{HT (left) and HH (right) axial $\hat{\mathbf{s}}$ strain (top plot) and $\hat{\mathbf{s}}$ component of polarisation $\mathbf{P}$ (bottom plot) as a function of position along the boundary normal $\hat{\mathbf{s}}$ implying their role as hole and electron sinks respectively. %
			Red and blue arrows represent the direction of $\hat{\mathbf{s}}$ polarisation components local to the wall (not to scale). A mechanism for enhancing charge carrier mobility and delaying recombination rate is proposed in the bottom panel.}
		\label{fig:locpot}
	\end{figure}

	We determined the formation energies of the twin boundaries to be $4.4\phantom{|}\mathrm{mJ\phantom{|}m^{-2}}$ and $-1.2\phantom{|}\mathrm{mJ \phantom{|}m^{-2}}$ for the HT and HH walls respectively \cite{wall_energy}. %
	The negative HH formation energy was found to arise from the octahedral tilt pattern across the wall. %
	Between both single antiphase tilt I4/mcm $a^0 a^0 c^-$ domains, there is an intermediary structure characterised by two antiphase axes of different tilt amplitudes (\textit{cf.}\ Fig.\ \ref{fig:structtiltpol}). %
	This phase is of space group $ C2/m $ with Glazer notation $a^{0}b^{-}c^{-}$. %
	Unusually, the $C2/m$ phase was found to be more stable than $I4/mcm$ in our PBEsol calculations, which correspond to the limit of 0~K. %
	We arrived at the same conclusion using the PBE functional and for different APbX\textsubscript{3} chemistries where A = Fr, Rb and X = I, F. %
	Nonetheless, these are small formation energies. %
	Hence, as has been observed experimentally \cite{RothmannLiZhuEtAl2017, LiuEtAl2018feTwinAPL, LiuEtAl2018feTwinNM}, we expect these walls to appear frequently. %

	The wall thicknesses were extracted from the change in octahedral tilts across the wall. %
	Figure \ref{fig:structtiltpol} shows plots of the magnitude of the cubic tilt vectors $\boldsymbol{\phi}$ with respect to position along the $\hat{\mathbf{s}}$ direction. %
	Fitting a hyperbolic tangent function $\sim \phi_0 \tanh(x/\delta)$ \cite{ITD_C3}, where $\phi_0$, $x$ and $\delta$ correspond to the bulk tilt amplitude, position along $\hat{\mathbf{s}}$ and the fitting parameter respectively, yields thicknesses of $2\delta \sim 1\phantom{|}\mathrm{nm}$ for both walls, indicating that they are thin. %
	For comparison, in the prototypical $I4/mcm$ oxide perovskite SrTiO\textsubscript{3}, the same wall geometries were calculated to have much larger thicknesses of $7\text{-}8\phantom{|}\text{nm}$ \cite{SchiaffinoStengel2017}. %
	Our results imply that any emergent phenomena at the walls could be highly localised. %
	Furthermore, these boundaries are likely to be relatively immobile in response to an external stress \cite{tagantsev}, which would help prevent wall annihilation. 
	
	We found both ferroelastic twins to be polar; a result that has not been discussed in their observation \cite{RothmannLiZhuEtAl2017, LiuEtAl2018feTwinAPL, LiuEtAl2018feTwinNM, LiuEtAlTwinRev2018, Rohm2017, Strelcove2017, MacDonald2017, Hermes2016}. %
	The layer-by-layer polarisation $\mathbf{P}$ was determined using Born effective charges computed in the centrosymmetric $Pm\bar{3}m$ structure \cite{BEC_formula}. %
	The $\hat{\mathbf{s}}\times\hat{\mathbf{r}}$ and $\hat{\mathbf{s}}$ components were antipolar and hence contributed no net polarisation across the wall. %
	Conversely, the $\hat{\mathbf{r}}$ component of $\mathbf{P}$ is plotted in Fig.\ \ref{fig:structtiltpol}, yielding a relatively large net polarisation peaking at $\sim 1.8 \text{ and } 6\phantom{|}\mu\mathrm{C\phantom{|}cm^{-2}}$ in the HT and HH walls respectively. %
	The same domain wall geometries in SrTiO\textsubscript{3} have been calculated to yield polarisations up to $0.2\phantom{|}\mu\mathrm{C\phantom{|}cm^{-2}}$ \cite{SchiaffinoStengel2017}. %
	We found this to be signficantly driven by Cs-cation off-centering in the HH wall, as opposed to that in the HT wall where the main contrbiution is from the larger Pb-cation Born effective charge. %
	In the hybrid organic-inorganic system, we might expect the dipole moment of MA to enhance this contribution to the polarisation. %
	
	Furthermore, we note that ferroelastic boundaries are necessarily polar \cite{ITD_C3}. %
	One may understand this as the juxtaposition of strain sates from both domains introducing a strain gradient across the wall. %
	This necessarily induces a flexoelectric effect (a coupling between strain gradient and electric polarisation) that breaks inversion symmetry \cite{SchiaffinoStengel2017}. %
	One may estimate this coupling to be particularly pronounced in these walls as shown in Fig.\ \ref{fig:structtiltpol}b) by the large gradient in shear strain. %
	It can be shown that the emergent polarisation at the wall is in-plane, corresponding to the $\hat{\mathbf{r}}$ direction in these HH and HT twins as confirmed by our results. %
	
	Although these walls are polar, this is only a necessary but not sufficient condition for ferroelectricity. %
	To shed some light on this issue, we note that Schiaffino and Stengel have derived a Landau-like expansion of the potential energy landscape for the same wall geometries \cite{SchiaffinoStengel2017}. %
	They identified three `improper' terms that are linear in $\mathbf{P}$ and arise due to the walls' structure. %
	These terms describe couplings of $\mathbf{P}$ with strain gradients (`flexoelectric'), tilts and tilt gradients (`rotopolar') in addition to a term coupling $\mathbf{P}$ with anti-polar B-cation off-centerings and tilts. %
	These tilt, shear strain and Pb $\hat{\mathbf{s}}\times\hat{\mathbf{r}}$ displacement profiles are plotted in Fig.\ \ref{fig:structtiltpol}. %
	For an improper term to remain invariant under $\mathbf{P}$ reversal, the sign of precisely one of the other terms in the coupling must also change. %
	Due to the wall geometry, that appears unlikely to be possible and hence we would not expect this improper polarisation to be switchable. %
	However, from computing phonon modes of the intermediate two (anti-phase) tilt system at the wall, we found this phase to contain an unstable polar mode. %
	This indicates a double well polarisation for CsPbI\textsubscript{3}, even after normalisation with the competitive bi-quadratic coupling term from the tilts, and hence the possibility of an additional proper and switchable component to the polarisation. %
	It is possible that a combination of these proper and improper polarisations may allow for the walls to be ferroelectric through the presence of multiple metastable polar states. %
	Nonetheless, we note that the aforementioned flexoelectricity and the photoferroic effect could play an important role in the photovoltaic performance of HOIPs, irrespective of whether or not $\mathbf{P}$ is switchable \cite{YangEtAl2018}. %
	The notion that strain and strain gradients bring about such effects in MAPI has been hypothesised \cite{TsaiEtAl2018,jones2018localarXiv} and, in addition to the characterisation of these twins, Liu \textit{et al}.\ \cite{LiuEtAl2018feTwinNM, LiuEtAl2018feTwinAPL} have imaged ionic segregation across the twin boundary which they attribute to the variation of strain. %
	
	Figure \ref{fig:locpot} shows an estimate for axial strain perpendicular to the wall \cite{layer_axial_strain} and the out-of-plane $\hat{\mathbf{s}}$ components of $\mathbf{P}$ as a function of distance along the $\hat{\mathbf{s}}$ direction. %
	The character of these polarisation components at the interface suggests that the HT and HH walls act as hole and electron sinks respectively. %
	We propose that this implies the interfaces to be pathways that could enhance electron/hole mobility. %
	Upon photoexcitation, an electron-hole pair may be separated into walls that act as their respective sinks, confining the charge carriers to the boundary and delaying recombination. %
	The in-plane wall polarisations then would enhance electron/hole mobility along the walls, similar to the ferroelectric highways depicted by Frost \textit{et al}.\ \cite{FrostEtAl2014}. %
	We have considered configurations where adjacent walls have parallel polarisation although antiparallel configurations are possible \cite{SchiaffinoStengel2017}. %
	In these ferroelastic twins, their experimental observation \cite{RothmannLiZhuEtAl2017} shows these walls to span over large distances and be a bulk phenomenon which may further support the proposed mechanism. %
	
	From the projected density of states we found the band gap at the walls and in bulk to be largely similar. %
	However, in general this may be an additional effect at different types of domain walls, which will be a point of future investigation. 

\section{Conclusions}
\label{sec:conc}

	In this article, we have presented an \textit{ab initio} study of ferroelastic $90^\circ$ CsPbI\textsubscript{3} twins as a model for the same defects observed in MAPI. %
	Both types of twins, HH and HT, were found to have a low formation energy, small thickness and a sizeable in-plane polarisation of $ \sim 6 \phantom{|} \mu\mathrm{C \phantom{|} cm^{-2}} $ despite separating non-polar domains. %
	Potential avenues for enhancing photovoltaic performance via domain wall engineering have been discussed with a mechanism proposed for delaying recombination rates and enhancing charge carrier mobility. %
	We hope to motivate further investigation into these structures and their role in the performance of hybrid organic-inorganic halide perovskite solar cells. %

\begin{acknowledgments}
 We are grateful for the computational resources provided by the Imperial College Research Computing Service and the UK Materials and Molecular Modelling Hub, which is partially funded by EPSRC (EP/P020194/1). %
 This work has been supported by the EPSRC Centre for Doctoral Training on Theory and Simulation of Materials (TSM-CDT, grant ref. EP/L015579/1). %
 J.\'I. thanks the support of the Luxembourg National Research Fund (Grant
 FNR/C15/MS/10458889 NEWALLS). %
 Structural schematics in Fig.\ \ref{fig:CsPbIi4mcm} \& \ref{fig:structtiltpol} were generated using the visualisation software VESTA 3 \cite{MommaVESTA}. %
\end{acknowledgments}

\bibliography{cspbi_paper}

\begin{thebibliography}{53}%
\makeatletter
\providecommand \@ifxundefined [1]{%
 \@ifx{#1\undefined}
}%
\providecommand \@ifnum [1]{%
 \ifnum #1\expandafter \@firstoftwo
 \else \expandafter \@secondoftwo
 \fi
}%
\providecommand \@ifx [1]{%
 \ifx #1\expandafter \@firstoftwo
 \else \expandafter \@secondoftwo
 \fi
}%
\providecommand \natexlab [1]{#1}%
\providecommand \enquote  [1]{``#1''}%
\providecommand \bibnamefont  [1]{#1}%
\providecommand \bibfnamefont [1]{#1}%
\providecommand \citenamefont [1]{#1}%
\providecommand \href@noop [0]{\@secondoftwo}%
\providecommand \href [0]{\begingroup \@sanitize@url \@href}%
\providecommand \@href[1]{\@@startlink{#1}\@@href}%
\providecommand \@@href[1]{\endgroup#1\@@endlink}%
\providecommand \@sanitize@url [0]{\catcode `\\12\catcode `\$12\catcode
  `\&12\catcode `\#12\catcode `\^12\catcode `\_12\catcode `\%12\relax}%
\providecommand \@@startlink[1]{}%
\providecommand \@@endlink[0]{}%
\providecommand \url  [0]{\begingroup\@sanitize@url \@url }%
\providecommand \@url [1]{\endgroup\@href {#1}{\urlprefix }}%
\providecommand \urlprefix  [0]{URL }%
\providecommand \Eprint [0]{\href }%
\providecommand \doibase [0]{http://dx.doi.org/}%
\providecommand \selectlanguage [0]{\@gobble}%
\providecommand \bibinfo  [0]{\@secondoftwo}%
\providecommand \bibfield  [0]{\@secondoftwo}%
\providecommand \translation [1]{[#1]}%
\providecommand \BibitemOpen [0]{}%
\providecommand \bibitemStop [0]{}%
\providecommand \bibitemNoStop [0]{.\EOS\space}%
\providecommand \EOS [0]{\spacefactor3000\relax}%
\providecommand \BibitemShut  [1]{\csname bibitem#1\endcsname}%
\let\auto@bib@innerbib\@empty
\bibitem [{\citenamefont {Rothmann}\ \emph {et~al.}(2017)\citenamefont
  {Rothmann}, \citenamefont {Li}, \citenamefont {Zhu}, \citenamefont {Bach},
  \citenamefont {Spiccia}, \citenamefont {Etheridge},\ and\ \citenamefont
  {Cheng}}]{RothmannLiZhuEtAl2017}%
  \BibitemOpen
  \bibfield  {author} {\bibinfo {author} {\bibfnamefont {M.~U.}\ \bibnamefont
  {Rothmann}}, \bibinfo {author} {\bibfnamefont {W.}~\bibnamefont {Li}},
  \bibinfo {author} {\bibfnamefont {Y.}~\bibnamefont {Zhu}}, \bibinfo {author}
  {\bibfnamefont {U.}~\bibnamefont {Bach}}, \bibinfo {author} {\bibfnamefont
  {L.}~\bibnamefont {Spiccia}}, \bibinfo {author} {\bibfnamefont
  {J.}~\bibnamefont {Etheridge}}, \ and\ \bibinfo {author} {\bibfnamefont
  {Y.-B.}\ \bibnamefont {Cheng}},\ }\href@noop {} {\bibfield  {journal}
  {\bibinfo  {journal} {Nat. Commun.}\ }\textbf {\bibinfo {volume} {8}},\
  \bibinfo {pages} {14547} (\bibinfo {year} {2017})}\BibitemShut {NoStop}%
\bibitem [{\citenamefont {Kojima}\ \emph {et~al.}(2009)\citenamefont {Kojima},
  \citenamefont {Teshima}, \citenamefont {Shirai},\ and\ \citenamefont
  {Miyasaka}}]{KojimaTeshimaShiraiEtAl2009}%
  \BibitemOpen
  \bibfield  {author} {\bibinfo {author} {\bibfnamefont {A.}~\bibnamefont
  {Kojima}}, \bibinfo {author} {\bibfnamefont {K.}~\bibnamefont {Teshima}},
  \bibinfo {author} {\bibfnamefont {Y.}~\bibnamefont {Shirai}}, \ and\ \bibinfo
  {author} {\bibfnamefont {T.}~\bibnamefont {Miyasaka}},\ }\href@noop {}
  {\bibfield  {journal} {\bibinfo  {journal} {J. Am. Chem. Soc.}\ }\textbf
  {\bibinfo {volume} {131}},\ \bibinfo {pages} {6050} (\bibinfo {year}
  {2009})}\BibitemShut {NoStop}%
\bibitem [{NRE()}]{NRELchart}%
  \BibitemOpen
  \href@noop {} {\enquote {\bibinfo {title} {Best {Research Cell Efficiencies,
  NREL}},}\ }\bibinfo {howpublished} {\url{https://www.nrel.gov/pv/}},\
  \bibinfo {note} {{Accessed}: 3rd January 2019}\BibitemShut {NoStop}%
\bibitem [{\citenamefont {Bhatt}\ and\ \citenamefont
  {Lee}(2017)}]{Bhatt2017review}%
  \BibitemOpen
  \bibfield  {author} {\bibinfo {author} {\bibfnamefont {M.~D.}\ \bibnamefont
  {Bhatt}}\ and\ \bibinfo {author} {\bibfnamefont {J.~S.}\ \bibnamefont
  {Lee}},\ }\href@noop {} {\bibfield  {journal} {\bibinfo  {journal} {New J.
  Chem.}\ }\textbf {\bibinfo {volume} {41}},\ \bibinfo {pages} {10508}
  (\bibinfo {year} {2017})}\BibitemShut {NoStop}%
\bibitem [{\citenamefont {Huang}\ \emph {et~al.}(2017)\citenamefont {Huang},
  \citenamefont {Yuan}, \citenamefont {Shao},\ and\ \citenamefont
  {Yan}}]{Huang2017NatReview}%
  \BibitemOpen
  \bibfield  {author} {\bibinfo {author} {\bibfnamefont {J.}~\bibnamefont
  {Huang}}, \bibinfo {author} {\bibfnamefont {Y.}~\bibnamefont {Yuan}},
  \bibinfo {author} {\bibfnamefont {Y.}~\bibnamefont {Shao}}, \ and\ \bibinfo
  {author} {\bibfnamefont {Y.}~\bibnamefont {Yan}},\ }\href@noop {} {\bibfield
  {journal} {\bibinfo  {journal} {Nat. Rev. Mater.}\ }\textbf {\bibinfo
  {volume} {2}},\ \bibinfo {pages} {17042 EP } (\bibinfo {year}
  {2017})}\BibitemShut {NoStop}%
\bibitem [{\citenamefont {Fridkin}(1979)}]{fridkin1979photoferroelectrics}%
  \BibitemOpen
  \bibfield  {author} {\bibinfo {author} {\bibfnamefont {V.~M.}\ \bibnamefont
  {Fridkin}},\ }\href@noop {} {\emph {\bibinfo {title} {Photoferroelectrics
  Springer Series in Solid State Sciences}}},\ Vol.~\bibinfo {volume} {9}\
  (\bibinfo  {publisher} {Springer-Verlag, Berlin},\ \bibinfo {year}
  {1979})\BibitemShut {NoStop}%
\bibitem [{\citenamefont {Sturman}\ and\ \citenamefont
  {Fridkin}(1992)}]{sturman1992photovoltaic}%
  \BibitemOpen
  \bibfield  {author} {\bibinfo {author} {\bibfnamefont {P.~J.}\ \bibnamefont
  {Sturman}}\ and\ \bibinfo {author} {\bibfnamefont {V.~M.}\ \bibnamefont
  {Fridkin}},\ }\href@noop {} {\emph {\bibinfo {title} {Photovoltaic and
  Photo-refractive Effects in Noncentrosymmetric Materials}}},\ Vol.~\bibinfo
  {volume} {8}\ (\bibinfo  {publisher} {CRC Press},\ \bibinfo {year}
  {1992})\BibitemShut {NoStop}%
\bibitem [{\citenamefont {Butler}\ \emph {et~al.}(2015)\citenamefont {Butler},
  \citenamefont {Frost},\ and\ \citenamefont {Walsh}}]{ButlerEtAl2015}%
  \BibitemOpen
  \bibfield  {author} {\bibinfo {author} {\bibfnamefont {K.~T.}\ \bibnamefont
  {Butler}}, \bibinfo {author} {\bibfnamefont {J.~M.}\ \bibnamefont {Frost}}, \
  and\ \bibinfo {author} {\bibfnamefont {A.}~\bibnamefont {Walsh}},\
  }\href@noop {} {\bibfield  {journal} {\bibinfo  {journal} {Energy Environ.
  Sci.}\ }\textbf {\bibinfo {volume} {8}},\ \bibinfo {pages} {838} (\bibinfo
  {year} {2015})}\BibitemShut {NoStop}%
\bibitem [{\citenamefont {Frost}\ \emph {et~al.}(2014)\citenamefont {Frost},
  \citenamefont {Butler}, \citenamefont {Brivio}, \citenamefont {Hendon},
  \citenamefont {van Schilfgaarde},\ and\ \citenamefont
  {Walsh}}]{FrostEtAl2014}%
  \BibitemOpen
  \bibfield  {author} {\bibinfo {author} {\bibfnamefont {J.~M.}\ \bibnamefont
  {Frost}}, \bibinfo {author} {\bibfnamefont {K.~T.}\ \bibnamefont {Butler}},
  \bibinfo {author} {\bibfnamefont {F.}~\bibnamefont {Brivio}}, \bibinfo
  {author} {\bibfnamefont {C.~H.}\ \bibnamefont {Hendon}}, \bibinfo {author}
  {\bibfnamefont {M.}~\bibnamefont {van Schilfgaarde}}, \ and\ \bibinfo
  {author} {\bibfnamefont {A.}~\bibnamefont {Walsh}},\ }\href@noop {}
  {\bibfield  {journal} {\bibinfo  {journal} {Nano Lett.}\ }\textbf {\bibinfo
  {volume} {14}},\ \bibinfo {pages} {2584} (\bibinfo {year}
  {2014})}\BibitemShut {NoStop}%
\bibitem [{\citenamefont {Kutes}\ \emph {et~al.}(2014)\citenamefont {Kutes},
  \citenamefont {Ye}, \citenamefont {Zhou}, \citenamefont {Pang}, \citenamefont
  {Huey},\ and\ \citenamefont {Padture}}]{KutesEtAl2014}%
  \BibitemOpen
  \bibfield  {author} {\bibinfo {author} {\bibfnamefont {Y.}~\bibnamefont
  {Kutes}}, \bibinfo {author} {\bibfnamefont {L.}~\bibnamefont {Ye}}, \bibinfo
  {author} {\bibfnamefont {Y.}~\bibnamefont {Zhou}}, \bibinfo {author}
  {\bibfnamefont {S.}~\bibnamefont {Pang}}, \bibinfo {author} {\bibfnamefont
  {B.~D.}\ \bibnamefont {Huey}}, \ and\ \bibinfo {author} {\bibfnamefont
  {N.~P.}\ \bibnamefont {Padture}},\ }\href@noop {} {\bibfield  {journal}
  {\bibinfo  {journal} {J. Phys. Chem. Lett.}\ }\textbf {\bibinfo {volume}
  {5}},\ \bibinfo {pages} {3335} (\bibinfo {year} {2014})}\BibitemShut
  {NoStop}%
\bibitem [{\citenamefont {Fan}\ \emph {et~al.}(2015)\citenamefont {Fan},
  \citenamefont {Xiao}, \citenamefont {Sun}, \citenamefont {Chen},
  \citenamefont {Hu}, \citenamefont {Ouyang}, \citenamefont {Ong},
  \citenamefont {Zeng},\ and\ \citenamefont {Wang}}]{FanEtAl2015}%
  \BibitemOpen
  \bibfield  {author} {\bibinfo {author} {\bibfnamefont {Z.}~\bibnamefont
  {Fan}}, \bibinfo {author} {\bibfnamefont {J.}~\bibnamefont {Xiao}}, \bibinfo
  {author} {\bibfnamefont {K.}~\bibnamefont {Sun}}, \bibinfo {author}
  {\bibfnamefont {L.}~\bibnamefont {Chen}}, \bibinfo {author} {\bibfnamefont
  {Y.}~\bibnamefont {Hu}}, \bibinfo {author} {\bibfnamefont {J.}~\bibnamefont
  {Ouyang}}, \bibinfo {author} {\bibfnamefont {K.~P.}\ \bibnamefont {Ong}},
  \bibinfo {author} {\bibfnamefont {K.}~\bibnamefont {Zeng}}, \ and\ \bibinfo
  {author} {\bibfnamefont {J.}~\bibnamefont {Wang}},\ }\href@noop {} {\bibfield
   {journal} {\bibinfo  {journal} {J. Phys. Chem. Lett.}\ }\textbf {\bibinfo
  {volume} {6}},\ \bibinfo {pages} {1155} (\bibinfo {year} {2015})}\BibitemShut
  {NoStop}%
\bibitem [{\citenamefont {G}\ \emph {et~al.}(2016)\citenamefont {G},
  \citenamefont {Mahale}, \citenamefont {Kore}, \citenamefont {Mukherjee},
  \citenamefont {Pavan}, \citenamefont {De}, \citenamefont {Ghara},
  \citenamefont {Sundaresan}, \citenamefont {Pandey}, \citenamefont
  {Guru~Row},\ and\ \citenamefont {Sarma}}]{SharadaEtAl2016}%
  \BibitemOpen
  \bibfield  {author} {\bibinfo {author} {\bibfnamefont {S.}~\bibnamefont {G}},
  \bibinfo {author} {\bibfnamefont {P.}~\bibnamefont {Mahale}}, \bibinfo
  {author} {\bibfnamefont {B.~P.}\ \bibnamefont {Kore}}, \bibinfo {author}
  {\bibfnamefont {S.}~\bibnamefont {Mukherjee}}, \bibinfo {author}
  {\bibfnamefont {M.~S.}\ \bibnamefont {Pavan}}, \bibinfo {author}
  {\bibfnamefont {C.}~\bibnamefont {De}}, \bibinfo {author} {\bibfnamefont
  {S.}~\bibnamefont {Ghara}}, \bibinfo {author} {\bibfnamefont
  {A.}~\bibnamefont {Sundaresan}}, \bibinfo {author} {\bibfnamefont
  {A.}~\bibnamefont {Pandey}}, \bibinfo {author} {\bibfnamefont {T.~N.}\
  \bibnamefont {Guru~Row}}, \ and\ \bibinfo {author} {\bibfnamefont {D.~D.}\
  \bibnamefont {Sarma}},\ }\href@noop {} {\bibfield  {journal} {\bibinfo
  {journal} {J. Phys. Chem. Lett.}\ }\textbf {\bibinfo {volume} {7}},\ \bibinfo
  {pages} {2412} (\bibinfo {year} {2016})}\BibitemShut {NoStop}%
\bibitem [{\citenamefont {Kim}\ \emph {et~al.}(2016)\citenamefont {Kim},
  \citenamefont {Dang}, \citenamefont {Choi}, \citenamefont {Park},
  \citenamefont {Eom}, \citenamefont {Song}, \citenamefont {Seol},
  \citenamefont {Kim}, \citenamefont {Shin}, \citenamefont {Nah},\ and\
  \citenamefont {Yoon}}]{KimEtAl2016}%
  \BibitemOpen
  \bibfield  {author} {\bibinfo {author} {\bibfnamefont {Y.-J.}\ \bibnamefont
  {Kim}}, \bibinfo {author} {\bibfnamefont {T.-V.}\ \bibnamefont {Dang}},
  \bibinfo {author} {\bibfnamefont {H.-J.}\ \bibnamefont {Choi}}, \bibinfo
  {author} {\bibfnamefont {B.-J.}\ \bibnamefont {Park}}, \bibinfo {author}
  {\bibfnamefont {J.-H.}\ \bibnamefont {Eom}}, \bibinfo {author} {\bibfnamefont
  {H.-A.}\ \bibnamefont {Song}}, \bibinfo {author} {\bibfnamefont
  {D.}~\bibnamefont {Seol}}, \bibinfo {author} {\bibfnamefont {Y.}~\bibnamefont
  {Kim}}, \bibinfo {author} {\bibfnamefont {S.-H.}\ \bibnamefont {Shin}},
  \bibinfo {author} {\bibfnamefont {J.}~\bibnamefont {Nah}}, \ and\ \bibinfo
  {author} {\bibfnamefont {S.-G.}\ \bibnamefont {Yoon}},\ }\href@noop {}
  {\bibfield  {journal} {\bibinfo  {journal} {J. Mater. Chem. A}\ }\textbf
  {\bibinfo {volume} {4}},\ \bibinfo {pages} {756} (\bibinfo {year}
  {2016})}\BibitemShut {NoStop}%
\bibitem [{\citenamefont {Rakita}\ \emph {et~al.}(2017)\citenamefont {Rakita},
  \citenamefont {Bar-Elli}, \citenamefont {Meirzadeh}, \citenamefont {Kaslasi},
  \citenamefont {Peleg}, \citenamefont {Hodes}, \citenamefont {Lubomirsky},
  \citenamefont {Oron}, \citenamefont {Ehre},\ and\ \citenamefont
  {Cahen}}]{RakitaEtAl2017}%
  \BibitemOpen
  \bibfield  {author} {\bibinfo {author} {\bibfnamefont {Y.}~\bibnamefont
  {Rakita}}, \bibinfo {author} {\bibfnamefont {O.}~\bibnamefont {Bar-Elli}},
  \bibinfo {author} {\bibfnamefont {E.}~\bibnamefont {Meirzadeh}}, \bibinfo
  {author} {\bibfnamefont {H.}~\bibnamefont {Kaslasi}}, \bibinfo {author}
  {\bibfnamefont {Y.}~\bibnamefont {Peleg}}, \bibinfo {author} {\bibfnamefont
  {G.}~\bibnamefont {Hodes}}, \bibinfo {author} {\bibfnamefont
  {I.}~\bibnamefont {Lubomirsky}}, \bibinfo {author} {\bibfnamefont
  {D.}~\bibnamefont {Oron}}, \bibinfo {author} {\bibfnamefont {D.}~\bibnamefont
  {Ehre}}, \ and\ \bibinfo {author} {\bibfnamefont {D.}~\bibnamefont {Cahen}},\
  }\href@noop {} {\bibfield  {journal} {\bibinfo  {journal} {Proc. Natl. Acad.
  Sci. U. S. A.}\ }\textbf {\bibinfo {volume} {114}},\ \bibinfo {pages} {E5504}
  (\bibinfo {year} {2017})}\BibitemShut {NoStop}%
\bibitem [{\citenamefont {Hoque}\ \emph {et~al.}(2016)\citenamefont {Hoque},
  \citenamefont {Yang}, \citenamefont {Li}, \citenamefont {Islam},
  \citenamefont {Pan}, \citenamefont {Zhu},\ and\ \citenamefont
  {Fan}}]{HoqueEtAl2016}%
  \BibitemOpen
  \bibfield  {author} {\bibinfo {author} {\bibfnamefont {M.~N.~F.}\
  \bibnamefont {Hoque}}, \bibinfo {author} {\bibfnamefont {M.}~\bibnamefont
  {Yang}}, \bibinfo {author} {\bibfnamefont {Z.}~\bibnamefont {Li}}, \bibinfo
  {author} {\bibfnamefont {N.}~\bibnamefont {Islam}}, \bibinfo {author}
  {\bibfnamefont {X.}~\bibnamefont {Pan}}, \bibinfo {author} {\bibfnamefont
  {K.}~\bibnamefont {Zhu}}, \ and\ \bibinfo {author} {\bibfnamefont
  {Z.}~\bibnamefont {Fan}},\ }\href@noop {} {\bibfield  {journal} {\bibinfo
  {journal} {ACS Energy Lett.}\ }\textbf {\bibinfo {volume} {1}},\ \bibinfo
  {pages} {142} (\bibinfo {year} {2016})}\BibitemShut {NoStop}%
\bibitem [{\citenamefont {Whalley}\ \emph {et~al.}(2017)\citenamefont
  {Whalley}, \citenamefont {Frost}, \citenamefont {Jung},\ and\ \citenamefont
  {Walsh}}]{WhalleyEtAl2017}%
  \BibitemOpen
  \bibfield  {author} {\bibinfo {author} {\bibfnamefont {L.~D.}\ \bibnamefont
  {Whalley}}, \bibinfo {author} {\bibfnamefont {J.~M.}\ \bibnamefont {Frost}},
  \bibinfo {author} {\bibfnamefont {Y.-K.}\ \bibnamefont {Jung}}, \ and\
  \bibinfo {author} {\bibfnamefont {A.}~\bibnamefont {Walsh}},\ }\href@noop {}
  {\bibfield  {journal} {\bibinfo  {journal} {J. Chem. Phys.}\ }\textbf
  {\bibinfo {volume} {146}},\ \bibinfo {pages} {220901} (\bibinfo {year}
  {2017})}\BibitemShut {NoStop}%
\bibitem [{\citenamefont {Liu}\ \emph {et~al.}(2018{\natexlab{a}})\citenamefont
  {Liu}, \citenamefont {Liu}, \citenamefont {Wang}, \citenamefont {Wu},
  \citenamefont {Liu}, \citenamefont {Xiao}, \citenamefont {Chen},
  \citenamefont {Wang},\ and\ \citenamefont {Gong}}]{LiuEtAlTwinRev2018}%
  \BibitemOpen
  \bibfield  {author} {\bibinfo {author} {\bibfnamefont {W.}~\bibnamefont
  {Liu}}, \bibinfo {author} {\bibfnamefont {Y.}~\bibnamefont {Liu}}, \bibinfo
  {author} {\bibfnamefont {J.}~\bibnamefont {Wang}}, \bibinfo {author}
  {\bibfnamefont {C.}~\bibnamefont {Wu}}, \bibinfo {author} {\bibfnamefont
  {C.}~\bibnamefont {Liu}}, \bibinfo {author} {\bibfnamefont {L.}~\bibnamefont
  {Xiao}}, \bibinfo {author} {\bibfnamefont {Z.}~\bibnamefont {Chen}}, \bibinfo
  {author} {\bibfnamefont {S.}~\bibnamefont {Wang}}, \ and\ \bibinfo {author}
  {\bibfnamefont {Q.}~\bibnamefont {Gong}},\ }\href@noop {} {\bibfield
  {journal} {\bibinfo  {journal} {Crystals}\ }\textbf {\bibinfo {volume} {8}}
  (\bibinfo {year} {2018}{\natexlab{a}})}\BibitemShut {NoStop}%
\bibitem [{\citenamefont {Shan}\ and\ \citenamefont
  {Saidi}(2017)}]{ShanSaidi2017GB}%
  \BibitemOpen
  \bibfield  {author} {\bibinfo {author} {\bibfnamefont {W.}~\bibnamefont
  {Shan}}\ and\ \bibinfo {author} {\bibfnamefont {W.~A.}\ \bibnamefont
  {Saidi}},\ }\href@noop {} {\bibfield  {journal} {\bibinfo  {journal} {J.
  Phys. Chem. Lett.}\ }\textbf {\bibinfo {volume} {8}},\ \bibinfo {pages}
  {5935} (\bibinfo {year} {2017})}\BibitemShut {NoStop}%
\bibitem [{\citenamefont {Yin}\ \emph {et~al.}(2014)\citenamefont {Yin},
  \citenamefont {Shi},\ and\ \citenamefont {Yan}}]{WanJianEtAl2014GB}%
  \BibitemOpen
  \bibfield  {author} {\bibinfo {author} {\bibfnamefont {W.-J.}\ \bibnamefont
  {Yin}}, \bibinfo {author} {\bibfnamefont {T.}~\bibnamefont {Shi}}, \ and\
  \bibinfo {author} {\bibfnamefont {Y.}~\bibnamefont {Yan}},\ }\href@noop {}
  {\bibfield  {journal} {\bibinfo  {journal} {Adv. Mater.}\ }\textbf {\bibinfo
  {volume} {26}},\ \bibinfo {pages} {4653} (\bibinfo {year}
  {2014})}\BibitemShut {NoStop}%
\bibitem [{\citenamefont {McKenna}(2018)}]{McKenna2018faGB}%
  \BibitemOpen
  \bibfield  {author} {\bibinfo {author} {\bibfnamefont {K.~P.}\ \bibnamefont
  {McKenna}},\ }\href@noop {} {\bibfield  {journal} {\bibinfo  {journal} {ACS
  Energy Lett.}\ }\textbf {\bibinfo {volume} {3}},\ \bibinfo {pages} {2663}
  (\bibinfo {year} {2018})}\BibitemShut {NoStop}%
\bibitem [{\citenamefont {Guo}\ \emph {et~al.}(2017)\citenamefont {Guo},
  \citenamefont {Wang},\ and\ \citenamefont {Saidi}}]{GuoEtAl2017GB}%
  \BibitemOpen
  \bibfield  {author} {\bibinfo {author} {\bibfnamefont {Y.}~\bibnamefont
  {Guo}}, \bibinfo {author} {\bibfnamefont {Q.}~\bibnamefont {Wang}}, \ and\
  \bibinfo {author} {\bibfnamefont {W.~A.}\ \bibnamefont {Saidi}},\ }\href@noop
  {} {\bibfield  {journal} {\bibinfo  {journal} {J. Phys. Chem. C}\ }\textbf
  {\bibinfo {volume} {121}},\ \bibinfo {pages} {1715} (\bibinfo {year}
  {2017})}\BibitemShut {NoStop}%
\bibitem [{\citenamefont {Yin}\ \emph {et~al.}(2015)\citenamefont {Yin},
  \citenamefont {Chen}, \citenamefont {Shi}, \citenamefont {Wei},\ and\
  \citenamefont {Yan}}]{WanJianEtAl2015GB}%
  \BibitemOpen
  \bibfield  {author} {\bibinfo {author} {\bibfnamefont {W.-J.}\ \bibnamefont
  {Yin}}, \bibinfo {author} {\bibfnamefont {H.}~\bibnamefont {Chen}}, \bibinfo
  {author} {\bibfnamefont {T.}~\bibnamefont {Shi}}, \bibinfo {author}
  {\bibfnamefont {S.-H.}\ \bibnamefont {Wei}}, \ and\ \bibinfo {author}
  {\bibfnamefont {Y.}~\bibnamefont {Yan}},\ }\href@noop {} {\bibfield
  {journal} {\bibinfo  {journal} {Adv. Electron. Mater.}\ }\textbf {\bibinfo
  {volume} {1}},\ \bibinfo {pages} {1500044} (\bibinfo {year}
  {2015})}\BibitemShut {NoStop}%
\bibitem [{\citenamefont {Liu}\ \emph {et~al.}(2015)\citenamefont {Liu},
  \citenamefont {Zheng}, \citenamefont {Koocher}, \citenamefont {Takenaka},
  \citenamefont {Wang},\ and\ \citenamefont {Rappe}}]{LiuZhengKoocherEtAl2015}%
  \BibitemOpen
  \bibfield  {author} {\bibinfo {author} {\bibfnamefont {S.}~\bibnamefont
  {Liu}}, \bibinfo {author} {\bibfnamefont {F.}~\bibnamefont {Zheng}}, \bibinfo
  {author} {\bibfnamefont {N.~Z.}\ \bibnamefont {Koocher}}, \bibinfo {author}
  {\bibfnamefont {H.}~\bibnamefont {Takenaka}}, \bibinfo {author}
  {\bibfnamefont {F.}~\bibnamefont {Wang}}, \ and\ \bibinfo {author}
  {\bibfnamefont {A.~M.}\ \bibnamefont {Rappe}},\ }\href@noop {} {\bibfield
  {journal} {\bibinfo  {journal} {J. Phys. Chem. Lett.}\ }\textbf {\bibinfo
  {volume} {6}},\ \bibinfo {pages} {693} (\bibinfo {year} {2015})}\BibitemShut
  {NoStop}%
\bibitem [{\citenamefont {Chen}\ \emph {et~al.}()\citenamefont {Chen},
  \citenamefont {Paillard}, \citenamefont {Zhao}, \citenamefont
  {{\'{I}}{\~{n}}iguez}, \citenamefont {Yang},\ and\ \citenamefont
  {Bellaiche}}]{ChenEtAl2018FEDW}%
  \BibitemOpen
  \bibfield  {author} {\bibinfo {author} {\bibfnamefont {L.}~\bibnamefont
  {Chen}}, \bibinfo {author} {\bibfnamefont {C.}~\bibnamefont {Paillard}},
  \bibinfo {author} {\bibfnamefont {H.~J.}\ \bibnamefont {Zhao}}, \bibinfo
  {author} {\bibfnamefont {J.}~\bibnamefont {{\'{I}}{\~{n}}iguez}}, \bibinfo
  {author} {\bibfnamefont {Y.}~\bibnamefont {Yang}}, \ and\ \bibinfo {author}
  {\bibfnamefont {L.}~\bibnamefont {Bellaiche}},\ }\href@noop {} {\bibinfo
  {journal} {npj Comput. Mater.}\ ,\ \bibinfo {pages} {75}}\BibitemShut
  {NoStop}%
\bibitem [{\citenamefont {Bischak}\ \emph {et~al.}(2015)\citenamefont
  {Bischak}, \citenamefont {Sanehira}, \citenamefont {Precht}, \citenamefont
  {Luther},\ and\ \citenamefont {Ginsberg}}]{BischakEtAl2015GB}%
  \BibitemOpen
\bibfield  {journal} {  }\bibfield  {author} {\bibinfo {author} {\bibfnamefont
  {C.~G.}\ \bibnamefont {Bischak}}, \bibinfo {author} {\bibfnamefont {E.~M.}\
  \bibnamefont {Sanehira}}, \bibinfo {author} {\bibfnamefont {J.~T.}\
  \bibnamefont {Precht}}, \bibinfo {author} {\bibfnamefont {J.~M.}\
  \bibnamefont {Luther}}, \ and\ \bibinfo {author} {\bibfnamefont {N.~S.}\
  \bibnamefont {Ginsberg}},\ }\href@noop {} {\bibfield  {journal} {\bibinfo
  {journal} {Nano Lett.}\ }\textbf {\bibinfo {volume} {15}},\ \bibinfo {pages}
  {4799} (\bibinfo {year} {2015})}\BibitemShut {NoStop}%
\bibitem [{\citenamefont {Liu}\ \emph {et~al.}(2018{\natexlab{b}})\citenamefont
  {Liu}, \citenamefont {Collins}, \citenamefont {Belianinov}, \citenamefont
  {Neumayer}, \citenamefont {Ievlev}, \citenamefont {Ahmadi}, \citenamefont
  {Xiao}, \citenamefont {Retterer}, \citenamefont {Jesse}, \citenamefont
  {Kalinin}, \citenamefont {Hu},\ and\ \citenamefont
  {Ovchinnikova}}]{LiuEtAl2018feTwinAPL}%
  \BibitemOpen
  \bibfield  {author} {\bibinfo {author} {\bibfnamefont {Y.}~\bibnamefont
  {Liu}}, \bibinfo {author} {\bibfnamefont {L.}~\bibnamefont {Collins}},
  \bibinfo {author} {\bibfnamefont {A.}~\bibnamefont {Belianinov}}, \bibinfo
  {author} {\bibfnamefont {S.~M.}\ \bibnamefont {Neumayer}}, \bibinfo {author}
  {\bibfnamefont {A.~V.}\ \bibnamefont {Ievlev}}, \bibinfo {author}
  {\bibfnamefont {M.}~\bibnamefont {Ahmadi}}, \bibinfo {author} {\bibfnamefont
  {K.}~\bibnamefont {Xiao}}, \bibinfo {author} {\bibfnamefont {S.~T.}\
  \bibnamefont {Retterer}}, \bibinfo {author} {\bibfnamefont {S.}~\bibnamefont
  {Jesse}}, \bibinfo {author} {\bibfnamefont {S.~V.}\ \bibnamefont {Kalinin}},
  \bibinfo {author} {\bibfnamefont {B.}~\bibnamefont {Hu}}, \ and\ \bibinfo
  {author} {\bibfnamefont {O.~S.}\ \bibnamefont {Ovchinnikova}},\ }\href@noop
  {} {\bibfield  {journal} {\bibinfo  {journal} {Appl. Phys. Lett.}\ }\textbf
  {\bibinfo {volume} {113}},\ \bibinfo {pages} {072102} (\bibinfo {year}
  {2018}{\natexlab{b}})}\BibitemShut {NoStop}%
\bibitem [{\citenamefont {Liu}\ \emph {et~al.}(2018{\natexlab{c}})\citenamefont
  {Liu}, \citenamefont {Collins}, \citenamefont {Proksch}, \citenamefont {Kim},
  \citenamefont {Watson}, \citenamefont {Doughty}, \citenamefont {Calhoun},
  \citenamefont {Ahmadi}, \citenamefont {Ievlev}, \citenamefont {Jesse},
  \citenamefont {Retterer}, \citenamefont {Belianinov}, \citenamefont {Xiao},
  \citenamefont {Huang}, \citenamefont {Sumpter}, \citenamefont {Kalinin},
  \citenamefont {Hu},\ and\ \citenamefont
  {Ovchinnikova}}]{LiuEtAl2018feTwinNM}%
  \BibitemOpen
  \bibfield  {author} {\bibinfo {author} {\bibfnamefont {Y.}~\bibnamefont
  {Liu}}, \bibinfo {author} {\bibfnamefont {L.}~\bibnamefont {Collins}},
  \bibinfo {author} {\bibfnamefont {R.}~\bibnamefont {Proksch}}, \bibinfo
  {author} {\bibfnamefont {S.}~\bibnamefont {Kim}}, \bibinfo {author}
  {\bibfnamefont {B.~R.}\ \bibnamefont {Watson}}, \bibinfo {author}
  {\bibfnamefont {B.}~\bibnamefont {Doughty}}, \bibinfo {author} {\bibfnamefont
  {T.~R.}\ \bibnamefont {Calhoun}}, \bibinfo {author} {\bibfnamefont
  {M.}~\bibnamefont {Ahmadi}}, \bibinfo {author} {\bibfnamefont {A.~V.}\
  \bibnamefont {Ievlev}}, \bibinfo {author} {\bibfnamefont {S.}~\bibnamefont
  {Jesse}}, \bibinfo {author} {\bibfnamefont {S.~T.}\ \bibnamefont {Retterer}},
  \bibinfo {author} {\bibfnamefont {A.}~\bibnamefont {Belianinov}}, \bibinfo
  {author} {\bibfnamefont {K.}~\bibnamefont {Xiao}}, \bibinfo {author}
  {\bibfnamefont {J.}~\bibnamefont {Huang}}, \bibinfo {author} {\bibfnamefont
  {B.~G.}\ \bibnamefont {Sumpter}}, \bibinfo {author} {\bibfnamefont {S.~V.}\
  \bibnamefont {Kalinin}}, \bibinfo {author} {\bibfnamefont {B.}~\bibnamefont
  {Hu}}, \ and\ \bibinfo {author} {\bibfnamefont {O.~S.}\ \bibnamefont
  {Ovchinnikova}},\ }\href@noop {} {\bibfield  {journal} {\bibinfo  {journal}
  {Nat. Mater.}\ } (\bibinfo {year} {2018}{\natexlab{c}})}\BibitemShut
  {NoStop}%
\bibitem [{ITD()}]{ITD_C3}%
  \BibitemOpen
  \href@noop {} {\emph {\bibinfo {title} {International Tables for
  Crystallography}}},\ Vol.~\bibinfo {volume} {D},\ Chap.~\bibinfo {chapter}
  {3}, pp.\ \bibinfo {pages} {358--559}\BibitemShut {NoStop}%
\bibitem [{\citenamefont {Lee}\ \emph {et~al.}(2016)\citenamefont {Lee},
  \citenamefont {Bristowe}, \citenamefont {Lee}, \citenamefont {Lee},
  \citenamefont {Bristowe}, \citenamefont {Cheetham},\ and\ \citenamefont
  {Jang}}]{LeeEtAl2016}%
  \BibitemOpen
  \bibfield  {author} {\bibinfo {author} {\bibfnamefont {J.-H.}\ \bibnamefont
  {Lee}}, \bibinfo {author} {\bibfnamefont {N.~C.}\ \bibnamefont {Bristowe}},
  \bibinfo {author} {\bibfnamefont {J.~H.}\ \bibnamefont {Lee}}, \bibinfo
  {author} {\bibfnamefont {S.-H.}\ \bibnamefont {Lee}}, \bibinfo {author}
  {\bibfnamefont {P.~D.}\ \bibnamefont {Bristowe}}, \bibinfo {author}
  {\bibfnamefont {A.~K.}\ \bibnamefont {Cheetham}}, \ and\ \bibinfo {author}
  {\bibfnamefont {H.~M.}\ \bibnamefont {Jang}},\ }\href@noop {} {\bibfield
  {journal} {\bibinfo  {journal} {Chem. Mater.}\ }\textbf {\bibinfo {volume}
  {28}},\ \bibinfo {pages} {4259} (\bibinfo {year} {2016})}\BibitemShut
  {NoStop}%
\bibitem [{\citenamefont {Brivio}\ \emph {et~al.}(2013)\citenamefont {Brivio},
  \citenamefont {Walker},\ and\ \citenamefont {Walsh}}]{BrivioWalkerWalsh2013}%
  \BibitemOpen
  \bibfield  {author} {\bibinfo {author} {\bibfnamefont {F.}~\bibnamefont
  {Brivio}}, \bibinfo {author} {\bibfnamefont {A.~B.}\ \bibnamefont {Walker}},
  \ and\ \bibinfo {author} {\bibfnamefont {A.}~\bibnamefont {Walsh}},\
  }\href@noop {} {\bibfield  {journal} {\bibinfo  {journal} {APL Mater.}\
  }\textbf {\bibinfo {volume} {1}},\ \bibinfo {pages} {042111} (\bibinfo {year}
  {2013})}\BibitemShut {NoStop}%
\bibitem [{\citenamefont {Eperon}\ \emph {et~al.}(2015)\citenamefont {Eperon},
  \citenamefont {Paternò}, \citenamefont {Sutton}, \citenamefont {Zampetti},
  \citenamefont {Haghighirad}, \citenamefont {Cacialli},\ and\ \citenamefont
  {Snaith}}]{EperonPaternoSuttonEtAl2015}%
  \BibitemOpen
  \bibfield  {author} {\bibinfo {author} {\bibfnamefont {G.~E.}\ \bibnamefont
  {Eperon}}, \bibinfo {author} {\bibfnamefont {G.~M.}\ \bibnamefont
  {Paternò}}, \bibinfo {author} {\bibfnamefont {R.~J.}\ \bibnamefont
  {Sutton}}, \bibinfo {author} {\bibfnamefont {A.}~\bibnamefont {Zampetti}},
  \bibinfo {author} {\bibfnamefont {A.~A.}\ \bibnamefont {Haghighirad}},
  \bibinfo {author} {\bibfnamefont {F.}~\bibnamefont {Cacialli}}, \ and\
  \bibinfo {author} {\bibfnamefont {H.~J.}\ \bibnamefont {Snaith}},\
  }\href@noop {} {\bibfield  {journal} {\bibinfo  {journal} {J. Mater. Chem.
  A}\ }\textbf {\bibinfo {volume} {3}},\ \bibinfo {pages} {19688} (\bibinfo
  {year} {2015})}\BibitemShut {NoStop}%
\bibitem [{\citenamefont {Choi}\ \emph {et~al.}(2014)\citenamefont {Choi},
  \citenamefont {Jeong}, \citenamefont {Kim}, \citenamefont {Kim},
  \citenamefont {Walker}, \citenamefont {Kim},\ and\ \citenamefont
  {Kim}}]{ChoiJeongKimEtAl2014}%
  \BibitemOpen
  \bibfield  {author} {\bibinfo {author} {\bibfnamefont {H.}~\bibnamefont
  {Choi}}, \bibinfo {author} {\bibfnamefont {J.}~\bibnamefont {Jeong}},
  \bibinfo {author} {\bibfnamefont {H.-B.}\ \bibnamefont {Kim}}, \bibinfo
  {author} {\bibfnamefont {S.}~\bibnamefont {Kim}}, \bibinfo {author}
  {\bibfnamefont {B.}~\bibnamefont {Walker}}, \bibinfo {author} {\bibfnamefont
  {G.-H.}\ \bibnamefont {Kim}}, \ and\ \bibinfo {author} {\bibfnamefont
  {J.~Y.}\ \bibnamefont {Kim}},\ }\href@noop {} {\bibfield  {journal} {\bibinfo
   {journal} {Nano Energy}\ }\textbf {\bibinfo {volume} {7}},\ \bibinfo {pages}
  {80 } (\bibinfo {year} {2014})}\BibitemShut {NoStop}%
\bibitem [{\citenamefont {Weller}\ \emph {et~al.}(2015)\citenamefont {Weller},
  \citenamefont {Weber}, \citenamefont {Henry}, \citenamefont {Di~Pumpo},\ and\
  \citenamefont {Hansen}}]{WellerWeberHenryEtAl2015}%
  \BibitemOpen
  \bibfield  {author} {\bibinfo {author} {\bibfnamefont {M.~T.}\ \bibnamefont
  {Weller}}, \bibinfo {author} {\bibfnamefont {O.~J.}\ \bibnamefont {Weber}},
  \bibinfo {author} {\bibfnamefont {P.~F.}\ \bibnamefont {Henry}}, \bibinfo
  {author} {\bibfnamefont {A.~M.}\ \bibnamefont {Di~Pumpo}}, \ and\ \bibinfo
  {author} {\bibfnamefont {T.~C.}\ \bibnamefont {Hansen}},\ }\href@noop {}
  {\bibfield  {journal} {\bibinfo  {journal} {Chem. Commun.}\ }\textbf
  {\bibinfo {volume} {51}},\ \bibinfo {pages} {4180} (\bibinfo {year}
  {2015})}\BibitemShut {NoStop}%
\bibitem [{\citenamefont {Glazer}(1972)}]{Glazer1972}%
  \BibitemOpen
  \bibfield  {author} {\bibinfo {author} {\bibfnamefont {A.~M.}\ \bibnamefont
  {Glazer}},\ }\href@noop {} {\bibfield  {journal} {\bibinfo  {journal} {Acta
  Crystallogr., Sect. B: Struct. Sci., Cryst. Eng. Mater.}\ }\textbf {\bibinfo
  {volume} {28}},\ \bibinfo {pages} {3384} (\bibinfo {year}
  {1972})}\BibitemShut {NoStop}%
\bibitem [{\citenamefont {Schiaffino}\ and\ \citenamefont
  {Stengel}(2017)}]{SchiaffinoStengel2017}%
  \BibitemOpen
  \bibfield  {author} {\bibinfo {author} {\bibfnamefont {A.}~\bibnamefont
  {Schiaffino}}\ and\ \bibinfo {author} {\bibfnamefont {M.}~\bibnamefont
  {Stengel}},\ }\href@noop {} {\bibfield  {journal} {\bibinfo  {journal} {Phys.
  Rev. Lett.}\ }\textbf {\bibinfo {volume} {119}},\ \bibinfo {pages} {137601}
  (\bibinfo {year} {2017})}\BibitemShut {NoStop}%
\bibitem [{\citenamefont {Perdew}\ \emph {et~al.}(2008)\citenamefont {Perdew},
  \citenamefont {Ruzsinszky}, \citenamefont {Csonka}, \citenamefont {Vydrov},
  \citenamefont {Scuseria}, \citenamefont {Constantin}, \citenamefont {Zhou},\
  and\ \citenamefont {Burke}}]{Perdew2008PBEsol}%
  \BibitemOpen
  \bibfield  {author} {\bibinfo {author} {\bibfnamefont {J.~P.}\ \bibnamefont
  {Perdew}}, \bibinfo {author} {\bibfnamefont {A.}~\bibnamefont {Ruzsinszky}},
  \bibinfo {author} {\bibfnamefont {G.~I.}\ \bibnamefont {Csonka}}, \bibinfo
  {author} {\bibfnamefont {O.~A.}\ \bibnamefont {Vydrov}}, \bibinfo {author}
  {\bibfnamefont {G.~E.}\ \bibnamefont {Scuseria}}, \bibinfo {author}
  {\bibfnamefont {L.~A.}\ \bibnamefont {Constantin}}, \bibinfo {author}
  {\bibfnamefont {X.}~\bibnamefont {Zhou}}, \ and\ \bibinfo {author}
  {\bibfnamefont {K.}~\bibnamefont {Burke}},\ }\href@noop {} {\bibfield
  {journal} {\bibinfo  {journal} {Phys. Rev. Lett.}\ }\textbf {\bibinfo
  {volume} {100}},\ \bibinfo {pages} {136406} (\bibinfo {year}
  {2008})}\BibitemShut {NoStop}%
\bibitem [{\citenamefont {Bl\"ochl}(1994)}]{Bloechl1994}%
  \BibitemOpen
  \bibfield  {author} {\bibinfo {author} {\bibfnamefont {P.~E.}\ \bibnamefont
  {Bl\"ochl}},\ }\href@noop {} {\bibfield  {journal} {\bibinfo  {journal}
  {Phys. Rev. B}\ }\textbf {\bibinfo {volume} {50}},\ \bibinfo {pages} {17953}
  (\bibinfo {year} {1994})}\BibitemShut {NoStop}%
\bibitem [{\citenamefont {Kresse}\ and\ \citenamefont
  {Joubert}(1999)}]{KresseJoubert1999}%
  \BibitemOpen
  \bibfield  {author} {\bibinfo {author} {\bibfnamefont {G.}~\bibnamefont
  {Kresse}}\ and\ \bibinfo {author} {\bibfnamefont {D.}~\bibnamefont
  {Joubert}},\ }\href@noop {} {\bibfield  {journal} {\bibinfo  {journal} {Phys.
  Rev. B}\ }\textbf {\bibinfo {volume} {59}},\ \bibinfo {pages} {1758}
  (\bibinfo {year} {1999})}\BibitemShut {NoStop}%
\bibitem [{\citenamefont {Kresse}\ and\ \citenamefont
  {Hafner}(1993)}]{KresseHafner1993}%
  \BibitemOpen
  \bibfield  {author} {\bibinfo {author} {\bibfnamefont {G.}~\bibnamefont
  {Kresse}}\ and\ \bibinfo {author} {\bibfnamefont {J.}~\bibnamefont
  {Hafner}},\ }\href@noop {} {\bibfield  {journal} {\bibinfo  {journal} {Phys.
  Rev. B}\ }\textbf {\bibinfo {volume} {47}},\ \bibinfo {pages} {558} (\bibinfo
  {year} {1993})}\BibitemShut {NoStop}%
\bibitem [{\citenamefont {Kresse}\ and\ \citenamefont
  {Furthm\"{u}ller}(1996)}]{KresseFurth1996}%
  \BibitemOpen
  \bibfield  {author} {\bibinfo {author} {\bibfnamefont {G.}~\bibnamefont
  {Kresse}}\ and\ \bibinfo {author} {\bibfnamefont {J.}~\bibnamefont
  {Furthm\"{u}ller}},\ }\href@noop {} {\bibfield  {journal} {\bibinfo
  {journal} {Comput. Mater. Sci.}\ }\textbf {\bibinfo {volume} {6}},\ \bibinfo
  {pages} {15 } (\bibinfo {year} {1996})}\BibitemShut {NoStop}%
\bibitem [{\citenamefont {Monkhorst}\ and\ \citenamefont
  {Pack}(1976)}]{MonkhorstPack1976}%
  \BibitemOpen
  \bibfield  {author} {\bibinfo {author} {\bibfnamefont {H.~J.}\ \bibnamefont
  {Monkhorst}}\ and\ \bibinfo {author} {\bibfnamefont {J.~D.}\ \bibnamefont
  {Pack}},\ }\href@noop {} {\bibfield  {journal} {\bibinfo  {journal} {Phys.\
  Rev.\ B}\ }\textbf {\bibinfo {volume} {13}},\ \bibinfo {pages} {5188}
  (\bibinfo {year} {1976})}\BibitemShut {NoStop}%
\bibitem [{wal()}]{wall_energy}%
  \BibitemOpen
  \href@noop {} {\bibinfo  {journal} {We define the wall formation energy,
  $E_{\mathrm{W}}$ as \begin{equation*} E_{\mathrm{W}} = \frac{1}{2}
  \frac{E_{\mathrm{SC}} - 10E_{\mathrm{20 atom cell}}}{S}, \end{equation*}
  where $E_{\mathrm{SC}}$ and $ E_{\mathrm{20 atom cell}} $ are the 200 atom
  wall cell and 20 atom conventional cells energies respectively and $S$ is the
  wall area contained in the supercell. The factor of $ 1/2 $ accounts for the
  two walls in the supercell}\ }\BibitemShut {NoStop}%
\bibitem [{\citenamefont {Tagantsev}\ \emph {et~al.}(2010)\citenamefont
  {Tagantsev}, \citenamefont {Cross},\ and\ \citenamefont
  {Fousek}}]{tagantsev}%
  \BibitemOpen
\bibfield  {journal} {  }\bibfield  {author} {\bibinfo {author} {\bibfnamefont
  {A.}~\bibnamefont {Tagantsev}}, \bibinfo {author} {\bibfnamefont {L.~E.}\
  \bibnamefont {Cross}}, \ and\ \bibinfo {author} {\bibfnamefont
  {J.}~\bibnamefont {Fousek}},\ }\href@noop {} {\emph {\bibinfo {title}
  {{Domain in Ferroic Crystals and Thin Films}}}}\ (\bibinfo  {publisher}
  {Springer-Verlag New York},\ \bibinfo {year} {2010})\BibitemShut {NoStop}%
\bibitem [{\citenamefont {R\"{o}hm}\ \emph {et~al.}(2017)\citenamefont
  {R\"{o}hm}, \citenamefont {Leonhard}, \citenamefont {Hoffmann},\ and\
  \citenamefont {Colsmann}}]{Rohm2017}%
  \BibitemOpen
  \bibfield  {author} {\bibinfo {author} {\bibfnamefont {H.}~\bibnamefont
  {R\"{o}hm}}, \bibinfo {author} {\bibfnamefont {T.}~\bibnamefont {Leonhard}},
  \bibinfo {author} {\bibfnamefont {M.~J.}\ \bibnamefont {Hoffmann}}, \ and\
  \bibinfo {author} {\bibfnamefont {A.}~\bibnamefont {Colsmann}},\ }\href@noop
  {} {\bibfield  {journal} {\bibinfo  {journal} {Energy Environ. Sci.}\
  }\textbf {\bibinfo {volume} {10}},\ \bibinfo {pages} {950} (\bibinfo {year}
  {2017})}\BibitemShut {NoStop}%
\bibitem [{\citenamefont {Strelcov}\ \emph {et~al.}(2017)\citenamefont
  {Strelcov}, \citenamefont {Dong}, \citenamefont {Li}, \citenamefont {Chae},
  \citenamefont {Shao}, \citenamefont {Deng}, \citenamefont {Gruverman},
  \citenamefont {Huang},\ and\ \citenamefont {Centrone}}]{Strelcove2017}%
  \BibitemOpen
  \bibfield  {author} {\bibinfo {author} {\bibfnamefont {E.}~\bibnamefont
  {Strelcov}}, \bibinfo {author} {\bibfnamefont {Q.}~\bibnamefont {Dong}},
  \bibinfo {author} {\bibfnamefont {T.}~\bibnamefont {Li}}, \bibinfo {author}
  {\bibfnamefont {J.}~\bibnamefont {Chae}}, \bibinfo {author} {\bibfnamefont
  {Y.}~\bibnamefont {Shao}}, \bibinfo {author} {\bibfnamefont {Y.}~\bibnamefont
  {Deng}}, \bibinfo {author} {\bibfnamefont {A.}~\bibnamefont {Gruverman}},
  \bibinfo {author} {\bibfnamefont {J.}~\bibnamefont {Huang}}, \ and\ \bibinfo
  {author} {\bibfnamefont {A.}~\bibnamefont {Centrone}},\ }\href@noop {}
  {\bibfield  {journal} {\bibinfo  {journal} {Sci. Adv.}\ }\textbf {\bibinfo
  {volume} {3}} (\bibinfo {year} {2017})}\BibitemShut {NoStop}%
\bibitem [{\citenamefont {MacDonald}\ \emph {et~al.}(2017)\citenamefont
  {MacDonald}, \citenamefont {Heveran}, \citenamefont {Yang}, \citenamefont
  {Moore}, \citenamefont {Zhu}, \citenamefont {Ferguson}, \citenamefont
  {Killgore},\ and\ \citenamefont {DelRio}}]{MacDonald2017}%
  \BibitemOpen
  \bibfield  {author} {\bibinfo {author} {\bibfnamefont {G.~A.}\ \bibnamefont
  {MacDonald}}, \bibinfo {author} {\bibfnamefont {C.~M.}\ \bibnamefont
  {Heveran}}, \bibinfo {author} {\bibfnamefont {M.}~\bibnamefont {Yang}},
  \bibinfo {author} {\bibfnamefont {D.}~\bibnamefont {Moore}}, \bibinfo
  {author} {\bibfnamefont {K.}~\bibnamefont {Zhu}}, \bibinfo {author}
  {\bibfnamefont {V.~L.}\ \bibnamefont {Ferguson}}, \bibinfo {author}
  {\bibfnamefont {J.~P.}\ \bibnamefont {Killgore}}, \ and\ \bibinfo {author}
  {\bibfnamefont {F.~W.}\ \bibnamefont {DelRio}},\ }\href@noop {} {\bibfield
  {journal} {\bibinfo  {journal} {ACS Appl. Mater. Interfaces}\ }\textbf
  {\bibinfo {volume} {9}},\ \bibinfo {pages} {33565} (\bibinfo {year}
  {2017})}\BibitemShut {NoStop}%
\bibitem [{\citenamefont {Hermes}\ \emph {et~al.}(2016)\citenamefont {Hermes},
  \citenamefont {Bretschneider}, \citenamefont {Bergmann}, \citenamefont {Li},
  \citenamefont {Klasen}, \citenamefont {Mars}, \citenamefont {Tremel},
  \citenamefont {Laquai}, \citenamefont {Butt}, \citenamefont {Mezger},
  \citenamefont {Berger}, \citenamefont {Rodriguez},\ and\ \citenamefont
  {Weber}}]{Hermes2016}%
  \BibitemOpen
  \bibfield  {author} {\bibinfo {author} {\bibfnamefont {I.~M.}\ \bibnamefont
  {Hermes}}, \bibinfo {author} {\bibfnamefont {S.~A.}\ \bibnamefont
  {Bretschneider}}, \bibinfo {author} {\bibfnamefont {V.~W.}\ \bibnamefont
  {Bergmann}}, \bibinfo {author} {\bibfnamefont {D.}~\bibnamefont {Li}},
  \bibinfo {author} {\bibfnamefont {A.}~\bibnamefont {Klasen}}, \bibinfo
  {author} {\bibfnamefont {J.}~\bibnamefont {Mars}}, \bibinfo {author}
  {\bibfnamefont {W.}~\bibnamefont {Tremel}}, \bibinfo {author} {\bibfnamefont
  {F.}~\bibnamefont {Laquai}}, \bibinfo {author} {\bibfnamefont {H.-J.}\
  \bibnamefont {Butt}}, \bibinfo {author} {\bibfnamefont {M.}~\bibnamefont
  {Mezger}}, \bibinfo {author} {\bibfnamefont {R.}~\bibnamefont {Berger}},
  \bibinfo {author} {\bibfnamefont {B.~J.}\ \bibnamefont {Rodriguez}}, \ and\
  \bibinfo {author} {\bibfnamefont {S.~A.~L.}\ \bibnamefont {Weber}},\
  }\href@noop {} {\bibfield  {journal} {\bibinfo  {journal} {J. Phys. Chem. C}\
  }\textbf {\bibinfo {volume} {120}},\ \bibinfo {pages} {5724} (\bibinfo {year}
  {2016})}\BibitemShut {NoStop}%
\bibitem [{BEC()}]{BEC_formula}%
  \BibitemOpen
  \href@noop {} {\bibinfo  {journal} {We calculated the layer by layer
  polarisation $\mathbf{P}$ with Born Effective Charges such that
  \begin{equation*} \mathbf{P} =
  \frac{1}{\Omega}\sum_{i}{w_{i}\mathbf{Z}_{i}\mathbf{u}_{i}} , \end{equation*}
  where $\mathbf{Z}_{i}$ and $\mathbf{u}_{i}$ denote the Born Effective Charge
  and displacement from a reference high symmetry position of the $i$th ion.
  Weights $w_i$ scale the contributions of ions shared with neighbouring cells
  that lie on the edges or vertices of the volume $\Omega$ (\textit{e.g}.\ an
  ion lying on a face--center is shared between two cells and hence
  $w_{i}=1/2$). The sum runs over all atoms contained in the volume in which
  the polarisation is computed}\ }\BibitemShut {NoStop}%
\bibitem [{\citenamefont {Yang}\ \emph {et~al.}(2018)\citenamefont {Yang},
  \citenamefont {Kim},\ and\ \citenamefont {Alexe}}]{YangEtAl2018}%
  \BibitemOpen
\bibfield  {journal} {  }\bibfield  {author} {\bibinfo {author} {\bibfnamefont
  {M.-M.}\ \bibnamefont {Yang}}, \bibinfo {author} {\bibfnamefont {D.~J.}\
  \bibnamefont {Kim}}, \ and\ \bibinfo {author} {\bibfnamefont
  {M.}~\bibnamefont {Alexe}},\ }\href@noop {} {\bibfield  {journal} {\bibinfo
  {journal} {Science}\ } (\bibinfo {year} {2018})}\BibitemShut {NoStop}%
\bibitem [{\citenamefont {Tsai}\ \emph {et~al.}(2018)\citenamefont {Tsai},
  \citenamefont {Asadpour}, \citenamefont {Blancon}, \citenamefont {Stoumpos},
  \citenamefont {Durand}, \citenamefont {Strzalka}, \citenamefont {Chen},
  \citenamefont {Verduzco}, \citenamefont {Ajayan}, \citenamefont {Tretiak},
  \citenamefont {Even}, \citenamefont {Alam}, \citenamefont {Kanatzidis},
  \citenamefont {Nie},\ and\ \citenamefont {Mohite}}]{TsaiEtAl2018}%
  \BibitemOpen
  \bibfield  {author} {\bibinfo {author} {\bibfnamefont {H.}~\bibnamefont
  {Tsai}}, \bibinfo {author} {\bibfnamefont {R.}~\bibnamefont {Asadpour}},
  \bibinfo {author} {\bibfnamefont {J.-C.}\ \bibnamefont {Blancon}}, \bibinfo
  {author} {\bibfnamefont {C.~C.}\ \bibnamefont {Stoumpos}}, \bibinfo {author}
  {\bibfnamefont {O.}~\bibnamefont {Durand}}, \bibinfo {author} {\bibfnamefont
  {J.~W.}\ \bibnamefont {Strzalka}}, \bibinfo {author} {\bibfnamefont
  {B.}~\bibnamefont {Chen}}, \bibinfo {author} {\bibfnamefont {R.}~\bibnamefont
  {Verduzco}}, \bibinfo {author} {\bibfnamefont {P.~M.}\ \bibnamefont
  {Ajayan}}, \bibinfo {author} {\bibfnamefont {S.}~\bibnamefont {Tretiak}},
  \bibinfo {author} {\bibfnamefont {J.}~\bibnamefont {Even}}, \bibinfo {author}
  {\bibfnamefont {M.~A.}\ \bibnamefont {Alam}}, \bibinfo {author}
  {\bibfnamefont {M.~G.}\ \bibnamefont {Kanatzidis}}, \bibinfo {author}
  {\bibfnamefont {W.}~\bibnamefont {Nie}}, \ and\ \bibinfo {author}
  {\bibfnamefont {A.~D.}\ \bibnamefont {Mohite}},\ }\href@noop {} {\bibfield
  {journal} {\bibinfo  {journal} {Science}\ }\textbf {\bibinfo {volume}
  {360}},\ \bibinfo {pages} {67} (\bibinfo {year} {2018})}\BibitemShut
  {NoStop}%
\bibitem [{\citenamefont {Jones}\ \emph {et~al.}(2018)\citenamefont {Jones},
  \citenamefont {Osherov}, \citenamefont {Alsari}, \citenamefont {Sponseller},
  \citenamefont {Duck}, \citenamefont {Jung}, \citenamefont {Settens},
  \citenamefont {Niroui}, \citenamefont {Brenes}, \citenamefont {Stan} \emph
  {et~al.}}]{jones2018localarXiv}%
  \BibitemOpen
  \bibfield  {author} {\bibinfo {author} {\bibfnamefont {T.~W.}\ \bibnamefont
  {Jones}}, \bibinfo {author} {\bibfnamefont {A.}~\bibnamefont {Osherov}},
  \bibinfo {author} {\bibfnamefont {M.}~\bibnamefont {Alsari}}, \bibinfo
  {author} {\bibfnamefont {M.}~\bibnamefont {Sponseller}}, \bibinfo {author}
  {\bibfnamefont {B.~C.}\ \bibnamefont {Duck}}, \bibinfo {author}
  {\bibfnamefont {Y.-K.}\ \bibnamefont {Jung}}, \bibinfo {author}
  {\bibfnamefont {C.}~\bibnamefont {Settens}}, \bibinfo {author} {\bibfnamefont
  {F.}~\bibnamefont {Niroui}}, \bibinfo {author} {\bibfnamefont
  {R.}~\bibnamefont {Brenes}}, \bibinfo {author} {\bibfnamefont {C.~V.}\
  \bibnamefont {Stan}},  \emph {et~al.},\ }\href@noop {} {\bibfield  {journal}
  {\bibinfo  {journal} {arXiv preprint arXiv:1803.01192}\ } (\bibinfo {year}
  {2018})}\BibitemShut {NoStop}%
\bibitem [{lay()}]{layer_axial_strain}%
  \BibitemOpen
  \href@noop {} {\bibinfo  {journal} {Axial $\hat{\mathbf{s}}$ strain was
  estimated by the expansion/compression of perpendicular distances between
  Cs--Pb--I planes relative to such a distance in the wall super cell where the
  bulk properties were recovered. Layer widths were calculated from taking
  distances between average positions of Cs and Pb cations in a given plane;
  these approximately correspond to the width of the dotted blue box along
  $\hat{\mathbf{s}}$ in Fig.\ \ref{fig:structtiltpol}a)}\ }\BibitemShut
  {NoStop}%
\bibitem [{\citenamefont {Momma}\ and\ \citenamefont
  {Izumi}(2011)}]{MommaVESTA}%
  \BibitemOpen
\bibfield  {journal} {  }\bibfield  {author} {\bibinfo {author} {\bibfnamefont
  {K.}~\bibnamefont {Momma}}\ and\ \bibinfo {author} {\bibfnamefont
  {F.}~\bibnamefont {Izumi}},\ }\href@noop {} {\bibfield  {journal} {\bibinfo
  {journal} {J. Appl. Crystallogr.}\ }\textbf {\bibinfo {volume} {44}},\
  \bibinfo {pages} {1272} (\bibinfo {year} {2011})}\BibitemShut {NoStop}%
\end{thebibliography}%

\end{document}